\documentstyle[psfig]{article}

\begin{document}

\include{epsf}

\begin{center}
{\large \bf Structure of binary Bose-Einstein condensates} \\
\vspace{0.5cm} {Marek Trippenbach$^{1,2}$, Krzysztof G\'{o}ral$^3$,
Kazimierz Rz\c{a}\.{z}ewski$^3$, Boris Malomed$^4$ and Y.\ B.\
Band$^1$} \\ \vspace{0.3cm}
{\footnotesize $^{\,1}$ Departments of Chemistry and Physics,
Ben-Gurion University of the Negev, Beer-Sheva, Israel \, 84105 \\
$^{\,2}$ Institute of Experimental Physics, Optics Division, Warsaw
University, ul.~Ho\.{z}a 69, Warsaw 00-681, Poland \\
$^{\,3}$ Center for Theoretical Physics and College of Science, \\
Polish Academy of Sciences, Al.  Lotnik\'{o}w 32/64, Warsaw 02-668,
Poland \\
$^{\,4}$ Department of Interdisciplinary Studies, Faculty of
Engineering, Tel-Aviv University, Tel-Aviv, Israel \, 69978 \\}
\end{center}




\begin{abstract}
We identify all possible classes of solutions for two-component
Bose-Einstein condensates (BECs) within the Thomas-Fermi (TF)
approximation, and check these results against numerical simulations
of the coupled Gross-Pitaevskii equations (GPEs).  We find that they
can be divided into two general categories.  The first class contains
solutions with a region of overlap between the components.  The other
class consists of non-overlapping wavefunctions, and contains also
solutions that do not possess the symmetry of the trap.  The chemical
potential and average energy can be found for both classes within the
TF approximation by solving a set of coupled algebraic equations
representing the normalization conditions for each component.  A
ground state minimizing the energy (within both classes of the states)
is found for a given set of parameters characterizing the scattering
length and confining potential.  In the TF approximation, the ground
state always shares the symmetry of the trap.  However, a full
numerical solution of the coupled GPEs, incorporating the kinetic
energy of the BEC atoms, can sometimes select a broken-symmetry state
as the ground state of the system.  We also investigate effects of
finite-range interactions on the structure of the ground state.
\end{abstract}


\section{Introduction}

Phase transitions and coexistence of different phases in
multi-component systems are of great importance to many areas of
physics, chemistry and biology.  An ideal system to study these
phenomena is a multi-component dilute atomic gas Bose-Einstein
condensate (BEC) mixture at zero temperature, due to the simplicity of
its theoretical description.  The mean-field approximation provides an
excellent description of these systems.  Other multi-component systems
cannot be understood as well as these BEC mixtures, because their
density is generally much higher, and the delta-function
pseudopotential, used to describe interactions between atoms in BEC,
is not appropriate for them.  Instead, a true microscopic interaction
potential must be employed to adequately describe such systems, hence
modeling them is much harder.

Multi-component BECs have been extensively studied over the last few
years \cite{Ho}-\cite{Stenholm}.  These studies have been motivated by
experimental work performed by the JILA \cite{JILA} and MIT
\cite{Kurn} groups.  Many interesting effects have been experimentally
determined and theoretically predicted, including topological
properties of the ground and excited states, phase transitions and
symmetry breaking \cite{ssb}, effects produced by a phase difference
between components \cite{Eberly}, stability properties \cite{Law},
Josephson-type oscillations \cite{Cooper}, four-wave mixing \cite
{4WM}, and trapping of boson-fermion and fermion-fermion systems \cite
{Molmer}.  Nevertheless, many features of BEC mixtures remain to be
explored by theorists and experimentalists.

A large variety of different species can be used to produce mixtures
of condensed bosons.  Mixtures of two different elements, or of
different isotopes of the same element, or simply different hyperfine
states of the same atom \cite{JILA,Kurn} can be considered.
Simulating experimental results for BEC binary mixtures requires
knowledge of the scattering lengths of the atoms involved.  To the
extent that the values of the scattering lengths are known with
insufficient accuracy, a full classification of different states is
necessary within the range of possible values.  This is also necessary
in the context of tuning the scattering length, as can be done by
means changing the external magnetic field near Feshbach resonances
Ref.~\cite {prospects}.  Varying the scattering length, one can study
phase transitions to states that break the symmetry of the trapping
potential.  Such states are known in the literature, and they were
observed in the JILA experiment \cite{JILA}.

The classification of two-component condensates can also be carried
out in systems with interconversion of components (``chemical
reactions'' between them), as in the case of an atom-molecule
condensate, where the conserved quantity is the number of atoms plus
twice the number of molecules (the number of atoms and number of
molecules are not separately conserved).  A mathematical model of the
latter system can be formulated in terms of two coupled
Gross-Pitaevskii equations (GPEs), which contain, in addition to the
familiar cubic self- and cross-interaction nonlinear terms, quadratic
terms that account for the ``chemistry'' (i.e., the interconversion)
\cite{Drummond}.  In particular, an interesting prediction of the
model is that a ``soliton'' state, i.e., a stationary self-supported
condensate cloud (similar to ``light bullets'' in nonlinear optics
\cite{bullets}), may exist without any trapping potential present
\cite{Drummond}.

Many aspects of binary condensate mixtures have been treated in the
literature, using various (mostly numerical) methods in order to
predict results for various experimental setups (see, e.g., Refs.
\cite{Esryprl}, \cite{Pu} and \cite{Ohberg}).  Nevertheless, a general
classification of all the ground-state solutions is not yet available.
This is understandable in view of many control parameters present in
the models (three scattering lengths, particle numbers for both
components and characteristics of the trap).  A very general and
elegant, but not explicit, algorithm for determining ground-state
shapes has been proposed by Ho and Shenoy \cite{Ho}.  Here we start
with the same goal in mind, but also with the intention to provide a
maximally straightforward and analytic set of expressions for the
ground-state wavefunctions and energies of atomic gas BEC mixtures.

We use a computational method simulating the evolution of
two-component GPEs in imaginary-time \cite{Esryprl} in order to study
phase separation of components in BEC mixtures.  Results produced by
this method are compared with analytical predictions based upon the
Thomas-Fermi (TF) approximation applied to two-component GPEs.  We
analyze the changes in the structure of separated phase BEC mixtures
with the variation of the $s$-wave scattering lengths and atom
numbers.  The changes can be predicted and understood using a simple
TF approximation, which includes equations obtained from normalization
conditions for both components.  The TF picture is compared with
solutions obtained using numerically simulated GPE evolution in
imaginary time, which includes the kinetic energy of atoms neglected
in the TF approximation.  We find that, for the simple case of a
spherically symmetric harmonic trapping potential, many
spherically-symmetric phase-separated geometries are possible,
depending on the ratios of the self- and cross- $s$-wave scattering
lengths for atomic collisions.  In the TF approximation,
symmetry-broken phase-separated geometries (i.e., those whose symmetry
is lower than that of the trap) are always energetically higher in
energy than those with unbroken symmetry.  Nevertheless, numerical
simulations in imaginary-time show that a lower-symmetry state may be
the lowest energy eigenstate, and thus determine a ground state of the
system.  Our method may be generalized to include a finite-range
interaction between atoms.  In the last section of the paper we study,
by means of direct numerical simulations, how such interactions affect
the geometry and shape of the ground state.

\section{Mean-field description of two-component Bose-Einstein mixtures}

\label{MFBEC}

In the present work, we concentrate on stationary states of BEC
mixtures, (not their dynamics), therefore we start with the
time-independent coupled GPEs, written in the standard notation:
\begin{equation}
\left( -\mu _{1}-\frac{\hbar^{2}\nabla^{2}}{2m_{1}}+V_{1}({\bf r}
)+U_{11}|\psi _{1}({\bf r},t)|^{2}+U_{12}|\psi _{2}({\bf r}
,t)|^{2}\right) \psi _{1}({\bf r},t)=0\ , \label{GP1}
\end{equation}
\begin{equation}
\left( -\mu _{2}-\frac{\hbar ^{2}\nabla ^{2}}{2m_{2}}+V_{2}({\bf r}) +
U_{12}|\psi _{1}({\bf r},t)|^{2}+U_{22}|\psi _{2}({\bf r},t)|^{2}
\right) \psi _{2}({\bf r},t)=0\ .  \label{GP2}
\end{equation}
Here $\mu _{1,2}$ are chemical potentials of the two species, and
$V_{1,2}({\bf r})$ are two isotropic parabolic trapping potentials,
i.e.,
\begin{equation}
V_{j}({\bf r})=(m_{j}/2)\omega _{j}^{2}\,r^{2},\,\,j=1,2,  \label{quadratic}
\end{equation}
$\omega _{j}$ being the corresponding frequencies of harmonic
oscillations of a trapped particle$\,$.  Further, $U_{ij}\equiv \left(
4\pi \hbar ^{2}/m_{ij}\right) \,a_{ij}$ are atom-atom interaction
strengths, proportional to the $s$-wave scattering lengths $a_{11}$,
$a_{22}$, and $a_{12}$ for the $1+1$, $2+2$, and $1+2$ collisions,
respectively, where $1$ and $2$ numerate the components, and
$m_{ij}=m_{i}$ if $i=j$ and $m_{ij} = m_{1}m_{2}/\left(
m_{1}+m_{2}\right)$ if $i\neq j$.  For simplicity, in the numerical
calculations presented here we take $m_{1}=m_{2}\equiv m$, and assume
that the magnetic moments of atoms belonging to the different
components are equal, so that the corresponding trapping potentials
are equal too, $\omega _{1}=\omega _{2}\equiv \omega $, but this
condition as well as the spherical symmetry condition may be readily
relaxed by means of rescaling variables.  Furthermore, we assume that
the scattering lengths are real, i.e., we assume that collisions are
not lossy.  We also assume that all the scattering lengths for both
different and alike atoms are positive; otherwise, the classification
of the possible states becomes very cumbersome.

Our calculations were carried out, simulating the evolution in the
{\em time-dependent} GPEs in imaginary time \cite{Esryprl}, so
that to let the solution relax to the ground state.  The
computations used the split operator method with the fast Fourier
transform, similar to that used in Ref.  \cite {Trippenbach-nl}.
The chemical potentials are obtained by computing the net energy
(the sum of kinetic, potential, self- and cross- nonlinear
mean-field energies) for each component. We have chosen the
wavefunctions $\psi _{1}({\bf r},t)$ and $\psi _{2}({\bf r},t)$ to
be normalized to the number of particles in each component, so
that $\int |\psi _{i}({\bf r},t)|^{2}d^{3}{\bf r}=N_{i}$. Choosing
the symmetry of an initial configuration in the imaginary-time
simulations, a solution $\psi _{1,2}({\bf r})$ which minimizes the
total energy
\begin{equation}
{\cal E} =   \int d^{3}{\bf r}\, [\mu _{1}|\psi _{1}({\bf
r},t)|^{2} + \mu _{2}|\psi _{2}({\bf r},t)|^{2}  -
\frac{U_{11}}{2}\,|\psi _{1}({\bf r},t)|^{4} - \frac{U_{22}}{2}\,
|\psi _{2}({\bf r},t)|^{4} - U_{12} \, |\psi _{1}({\bf
r},t)|^{2}|\psi _{2}({\bf r},t)|^{2} ] \ , \label{cal_E}
\end{equation}

within the class of functions possessing this symmetry can be found.
The ground state of the two-component Hamiltonian is the one with the
smallest value of ${\cal E}$ for different symmetry classes.

\section{Thomas-Fermi Approximation}

\label{TFA}

The TF approximation can be used to describe a zero temperature
condensate in the cases when the trapping-potential and mean-field
nonlinear terms in GPEs are attractive and repulsive respectively, and
the number of atoms is large so that the mean-field energies are large
compared to the kinetic energy.  For the two-component system with
overlapping wavefunctions,
\begin{equation}
\ \left(
\begin{array}{c}
\mu _{1}-V_{1}({\bf r}) \\
\mu _{2}-V_{2}({\bf r})
\end{array}
\right) =\left(
\begin{array}{cc}
U_{11} & U_{12} \\
U_{21} & U_{22}
\end{array}
\right) \left(
\begin{array}{c}
|\psi _{1}({\bf r},t)|^{2} \\
|\psi _{2}({\bf r},t)|^{2}
\end{array}
\right) \,.\   \label{GPTF}
\end{equation}
These equations can be solved as a linear system of equations for
$|\psi _{1}({\bf r},t)|^{2}$ and $|\psi _{2}({\bf r},t)|^{2}$ in terms
of the chemical potentials $\mu _{1}$ and $\mu _{2}$ to obtain:
\begin{equation}
|\psi _{1}({\bf r})|^{2} = \frac{\left[ \mu _{1}U_{11}-\mu _{2}U_{12}
\right] -\left[ U_{22}-U_{12}\right] (m/2)\omega
^{2}r^{2}}{U_{11}U_{22}-U_{12}U_{12}}\ , \label{GPTFsol1}
\end{equation}
\begin{equation}
|\psi_{2}({\bf r})|^{2} = \frac{\left[ \mu _{2}U_{11}-\mu
_{1}U_{12} \right] -\left[ U_{11}-U_{12}\right] (m/2)\omega
^{2}r^{2}}{U_{11}U_{22}-U_{12}U_{12}}\,.  \label{GPTFsol2}
\end{equation}
The chemical potentials $\mu _{1}$ and $\mu _{2}$ are determined from
the normalization conditions, $N_{i}=\int d^{D}x\,|\psi _{i}({\bf
r},t)|^{2}$, where $D$ is the dimension.  We shall plot examples for
$D=1$, but our numerical method is valid for higher dimensions as
well, hence we derive all the formulas for the general case.

When phase separation occurs and there are regions in the physical
space occupied by one component only, the TF approximation leads,
instead of Eqs.~(\ref{GPTF}), to the corresponding one-component TF
equations.  For example, if a phase with only species $1$ exists in a
particular region of space, an equation of the form
\begin{equation}
|\psi _{1}({\bf r},t)|^{2}=(\mu _{1}-V_{1}({\bf r}))/U_{11} \ ,
\label{single}
\end{equation}
is to be used in this region.  If another phase exists wherein the
species $1$ and $2$ are mixed, Eqs.~(\ref{GPTF}) are relevant for that
region.  The chemical potentials $\mu _{1}$ and $\mu _{2}$ must be
determined by setting the number of atoms of each type equal to the
integral of the corresponding density over the whole space.

Inspection of Eqs.~(\ref{GPTF}) suggests then that all the
solutions can be classified according the signs of three
parameters: $\det U\equiv U_{11}U_{22}-\left( U_{12}\right) ^{2}$,
and $\alpha _{j}\equiv U_{jj}-U_{12}$.  In particular, $\alpha
_{1,2}$ determine signs of the curvature (coefficients in front of
$r^{2}$) of the {\em effective} quadratic potentials for the two
components in Eqs.~(\ref{GPTFsol1}) and (\ref{GPTFsol2}).
Qualitatively different types of possible states with overlapping
wave functions (i.e., disregarding regions where only one of the
species is present) identified by the TF analysis are defined in
Table 1.

\begin{table}[tbp]
\centering \
\begin{tabular}{|c|c|c|c|}
\hline
& $U_{11}U_{22}-U_{12}U_{12}$ & $U_{11}-U_{12}$ & $
U_{22}-U_{12}$ \\
& $\equiv \det U$ & $\equiv \alpha _{1}$ & $\equiv \alpha _{2}$ \\
\hline\hline
Case 1 (Type A) & positive & positive & positive \\ \hline
Case 2 (Type B) & negative & negative & negative \\ \hline
Case 3 (Type A) & positive & negative & positive \\ \hline
Case 4 (Type A) & positive & positive & negative \\ \hline
Case 5 (Type B) & negative & positive & negative \\ \hline
Case 6 (Type B) & negative & negative & positive \\ \hline
Not possible & positive & negative & negative \\ \hline
Not possible & negative & positive & positive \\ \hline
\end{tabular}
\caption{Classification of the Thomas-Fermi forms.}
\end{table}

Let us focus on those cases when kinetic energy contribution does
not change the general structure of the solution, but only
generates narrow transient layers on the scale of the
corresponding healing length, as in the single-component case when
the TF approximation is valid (thus we consider large-size
condensates.  The analysis will include the TF configurations with
and without the overlap of the two different condensate
wavefunctions.  As already mentioned, in the most cases both
wavefunctions {\it do not overlap everywhere}, i.e., there is a
region where only one wavefunction is different from zero.
Consequently, search for the lowest-energy state of the mixture
cannot rely solely on Eqs.~(\ref{GPTFsol1})-(\ref{GPTFsol2})
obtained in the assumption that the overlapping takes place
everywhere.

Combining the cases represented in Table 1 and single-wavefunction
solutions within the TF approximation, we distinguish two general
types of solutions: unseparated ones, having an overlap region
where both wavefunctions are nonzero, and separated solutions
which do not contain any overlap region.  In the latter case, we
shall see from the analysis of a full GP equation (with kinetic
energy included) and also a {\em nonlocal} version of the
two-species model, with a finite range of the interatomic
interactions, that it is necessary to further distinguish between
weak ($U_{11}U_{22}\leq \left( U_{12}\right)^{2}$) and strong
($U_{11}U_{22}\ll \left( U_{12}\right)^{2}$) separation \cite{Ao}.

\subsection{Partially Overlapping Wavefunctions $(U_{11}U_{22}-\left(
U_{12}\right)^{2}>0)$} \label{over-lap}

\begin{figure}[ht]
\centerline{\epsfxsize=4.25in\epsfbox{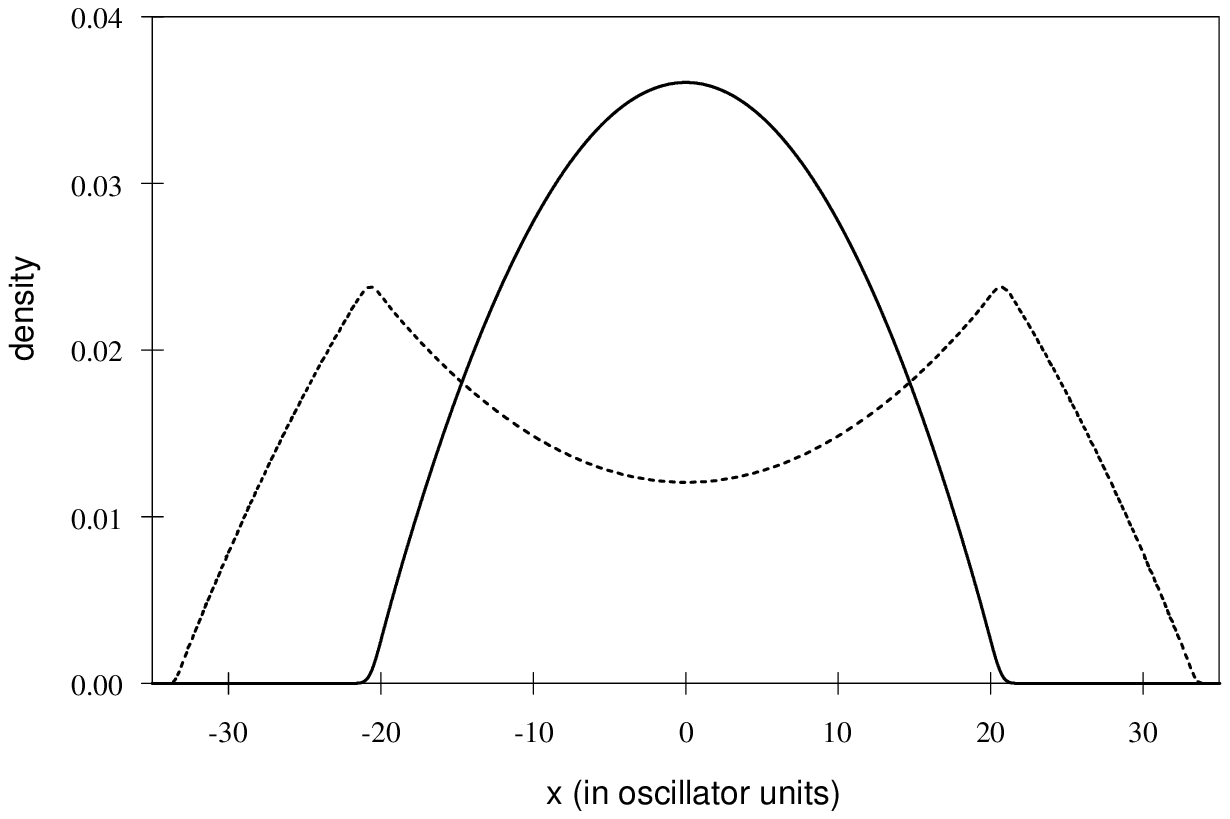}}
\centerline{\epsfxsize=4.25in\epsfbox{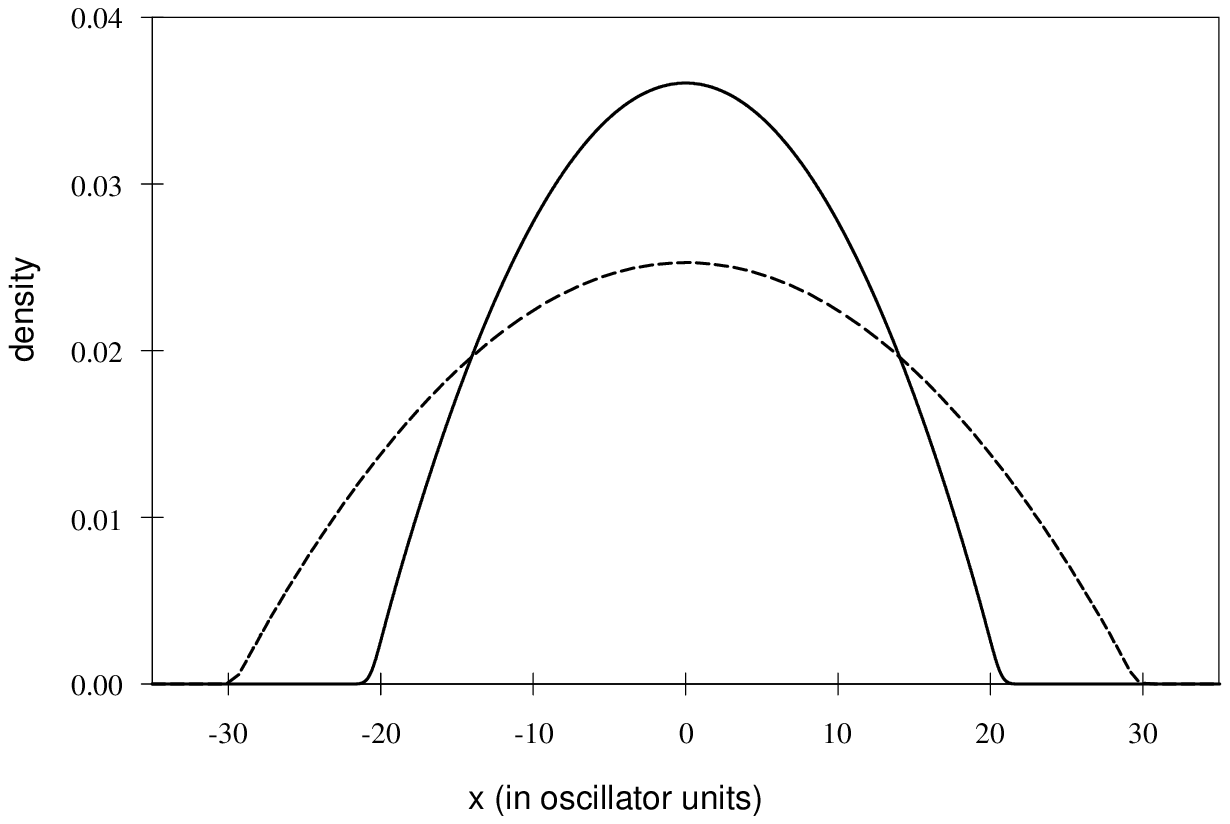}}
\caption{Two types of solutions with partial overlap.  Plotted are
densities for the first and the second component as solid and dashed
curves.  Panels (a) and (b) depict have $U_{11}:U_{12}:U_{22} =
1.2:0.9:0.8$ and $U_{11}:U_{12}:U_{22} = 1.1:0.9:1$ respectively.  In
both panels $N_{1}=N_{2}$.  The oscillator length unit is given by
$\sqrt{\hbar/m \omega}$.  Density distributions in all figures are 
normalized to unity.}
\label{overlap}
\end{figure}

In the case $\det U>0$ (cases 1, 3 and 4 in Table 1), we have checked
numerically that the minimum-energy solution is given by the
wavefunctions of the form shown in Fig.~\ref{overlap}.  Near the
origin, both wavefunctions coexist up to the point where one of them
vanishes.  Past this point, one wavefunction vanishes, while the other
one remains nonzero, following the single-component solution,
Eq.~(\ref{single}).  Fig.~\ref{overlap} shows two different cases that
are possible with the scenario described above.  In
Fig.~\ref{overlap}a, the two effective trapping potentials have the
same sign of their curvature, i.e., $\alpha _{1}\alpha _{2}>0$,\ in
the overlap region (this is case 1 in Table 1).  Fig.~\ref{overlap}b
presents another situation, when the two effective potentials have
opposite curvatures, $\alpha _{1}\alpha _{2}<0$ (these are cases 3 and
4 in Table 1).

\subsection{Separated Wavefunctions $(U_{11}U_{22}-\left(
U_{12}\right)^{2}<0)$ \label{separated}}

This category is represented by cases 2, 5 and 6 from Table 1.  The
simplest configuration is that with one wavefunction being different
from zero in the region around the origin and vanishing beyond a
separation radius, $R$, while the second component surrounds the first
one.  In this case, we can express $\mu _{1}$ and $\mu _{2}$ as
functions of $R$ and minimize the net free energy ${\cal E}$, in order
to find the lowest eigenstate of this type.  The normalization
conditions for the wavefunctions of the two components give a set of
relations between $R$, the chemical potentials $\mu _{1}$ and $\mu
_{2}$, and the number of atoms in each condensate:
\begin{equation}
N_{1}=\int_{0}^{R}\,d^{D}r\,\left[ \mu _{1}-V({\bf r})\right]
/U_{11}\,,
\label{CP1}
\end{equation}
\begin{equation}
N_{2}=\int_{R}^{R_{0}}\,d^{D}r\ \left[ \mu _{2}-V({\bf r})\right]
/U_{22}\,.
\label{CP2}
\end{equation}
Here $V({\bf r})$ is the binding potential (\ref{quadratic}), and
$R_{0}$ is a outer radius at which the wavefunction of the second
component vanishes in the FT approximation.  We first consider the 1D
case and then show how these considerations can be generalized to two
and three dimensions.

\subsubsection{One-Dimensional Case} \label{1D case}

To find the value of the radius $R$ minimizing ${\cal E}$, we solve
the set of the coupled equations (\ref{CP1}) and (\ref{CP2}) for $\mu
_{1}$ and $ \mu _{2}$.  The first equation can be solved directly to
yield $\mu _{1}$ as a function of $R$.  The second equation is more
complicated -- it can be solved analytically only in the 1D and 2D
cases, but not in 3D. In the 1D (3D) case, one needs to solve a third-
(fifth-) order algebraic equation to find $\mu _{2}$ as a function of
$R$.
The system of equations ({\ref{CP1}}) and ({\ref{CP2}}) in 1D
reduces to
\begin{equation} \label{EqR1}
2\,( \mu _{1}R - \frac{1}{6} m\omega^2 R^3) = U_{11}\, N_1 \,
\end{equation}
\begin{equation} \label{EqR2}
2\,( \frac{(2\mu_2)^{3/2}}{3\sqrt{m\omega^2}} - \mu _{2}R +
\frac{1}{6} m\omega^2 R^3) = U_{22}\, N_2 \ .
\end{equation}
In the case under consideration here, without spatial overlap of
components, the total energy is simply a sum of the average values of
harmonic potential and half of the nonlinear term in the GP equation
for each component state: $ {\cal E} =\frac{1}{2} \sum_i
<\psi_i|m\omega^2 x^2 + U_{ii}|\psi_i|^2/2|\psi_i>$.  If we substitute
the direct expression for the wavefunction in the TF approximation we
obtain:
\begin{equation}
{\cal E} = ( \frac{(2\mu_2)^{5/2}}{5\sqrt{m\omega^2}} - \mu _{2}^2
R + \frac{1}{20}(m\omega^2)^2R^5)/U_{22} + ( \mu _{1}^2R +
\frac{1}{20}(m\omega^2)^2R^5)/U_{11}
\end{equation}
and the chemical potentials can be found using
Eqs.~(\ref{EqR1})-(\ref{EqR2}).

\subsubsection{Generalization to Two and Three Dimensions}

A particularly simple form of these equations is obtained upon
introducing a TF radius $\left( R_{{\rm TF} }\right) _{1,2}$ of
the condensate in the corresponding dimension.  The TF radius is
defined as a radius of the{\it \ single} spherically-symmetric
condensate obtained in the TF approximation.  We can find an
explicit dependence between the chemical potentials of the
condensates and the separation radius of the two phases in 2D. In
this case, we again define the TF radii, which in 2D are equal to:
$\left( R_{{\rm TFi} }\right)^{4}=8N_iU_{ii}/(\pi m\omega ^{2})$.
In this case, we obtain the following set of equations for
chemical potentials and total energy:
\begin{equation}
\mu _{1}=\frac{m\omega ^{2}}{4R^{2}} \left[ \left( R_{{\rm
TF1}}\right)^{4}+R^{4}\right] \ ,
\end{equation}
\begin{equation}
\mu _{2}=\frac{m\omega ^{2}}{2} \left[ \left( R_{{\rm TF2}}\right)
^{2}+R^{2}\right] \ ,
\end{equation}
\begin{equation}
{\cal E} = \frac{1}{(m
\omega^2)^2}(\frac{2\mu_{1}^{2}r^{2}-\frac{1}{6}r^6}{R_{TF1}^4} +
\frac{\frac{8}{3}\mu_2^3 - 2\mu_2^2 r^2 +
\frac{1}{6}r^6}{R_{TF2}^4}) \ .
\end{equation}
where r is defined as $r = R \sqrt{m\omega^2}$.

In 3D we can define the TF radius is given by $\left( R_{{\rm
TFi}}\right){5}=(15U_{ii})/(2\pi m\omega^{2})$, and we can derive
a set of equations which can be solved for the chemical potentials
vs.~the separation radius:
\begin{equation}
\mu _{1}=\frac{m\omega ^{2}}{10R^{3}} \left[ 3R^{5}+2\left(
R_{{\rm TF1}}\right)^{5}\,\right] \,,
\end{equation}
\begin{equation}
4(\frac{2\mu_2}{m\omega^2})^{5/2}+3R^{5}-5\frac{2\mu_2}{m\omega^2}R^{3}-2\left(
R_{{\rm TF2}}\right)^{5}=0\,.
\end{equation}
The energy is given in terms of the separation radius R by:
\begin{equation}
{\cal E} = \frac{1}{({m\omega^2})^{5/2}} \left(
\frac{\frac{5}{2}\mu_1^2r^3 - \frac{15}{56}r^7}{R_{TF1}^5} +
\frac{\frac{5(2\mu_2)^{7/2}}{14} - \frac{5}{2}\mu_2^2r^3 +
\frac{15}{56}r^7}{R_{TF2}^5 } \right)
\end{equation}
where $r = R\sqrt{m\omega^2}$.

\subsubsection{Symmetry Breaking Solutions}

\begin{figure}[ht]
\centerline{\epsfxsize=4.25in\epsfbox{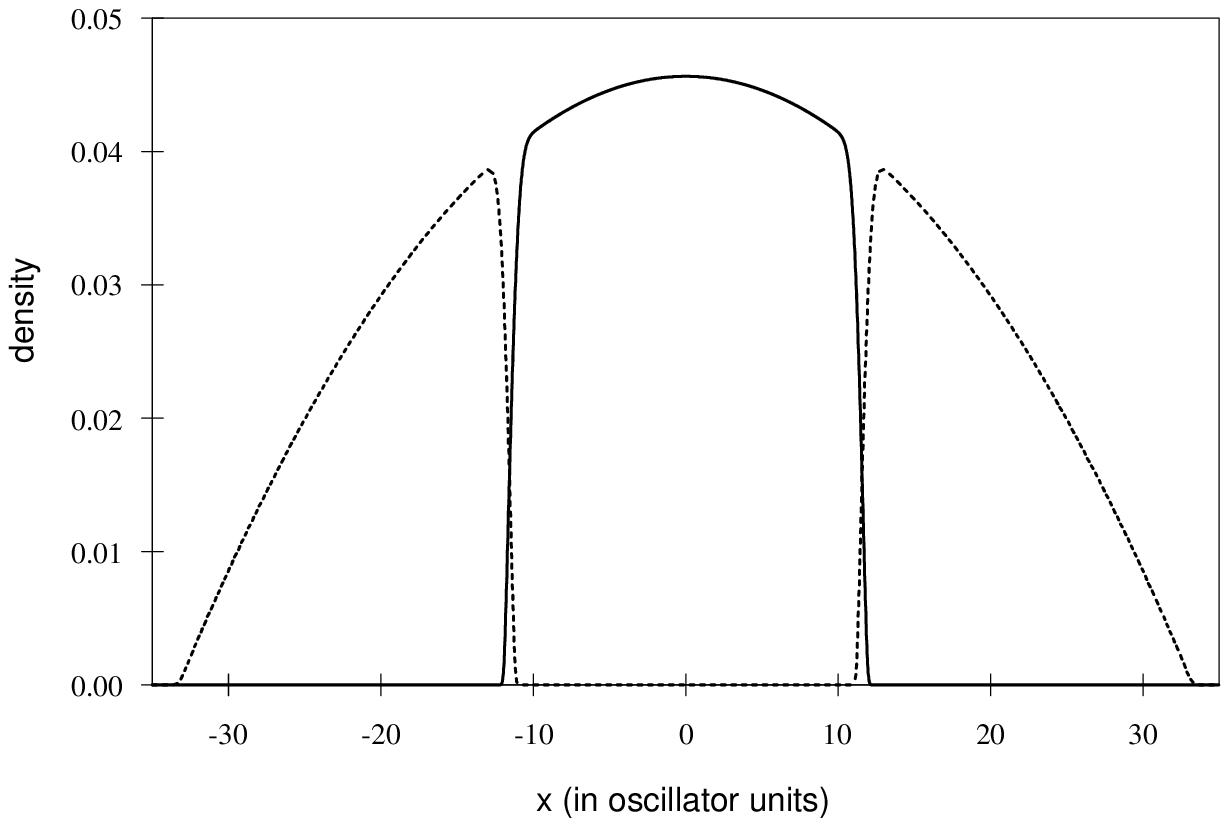}}
\centerline{\epsfxsize=4.25in\epsfbox{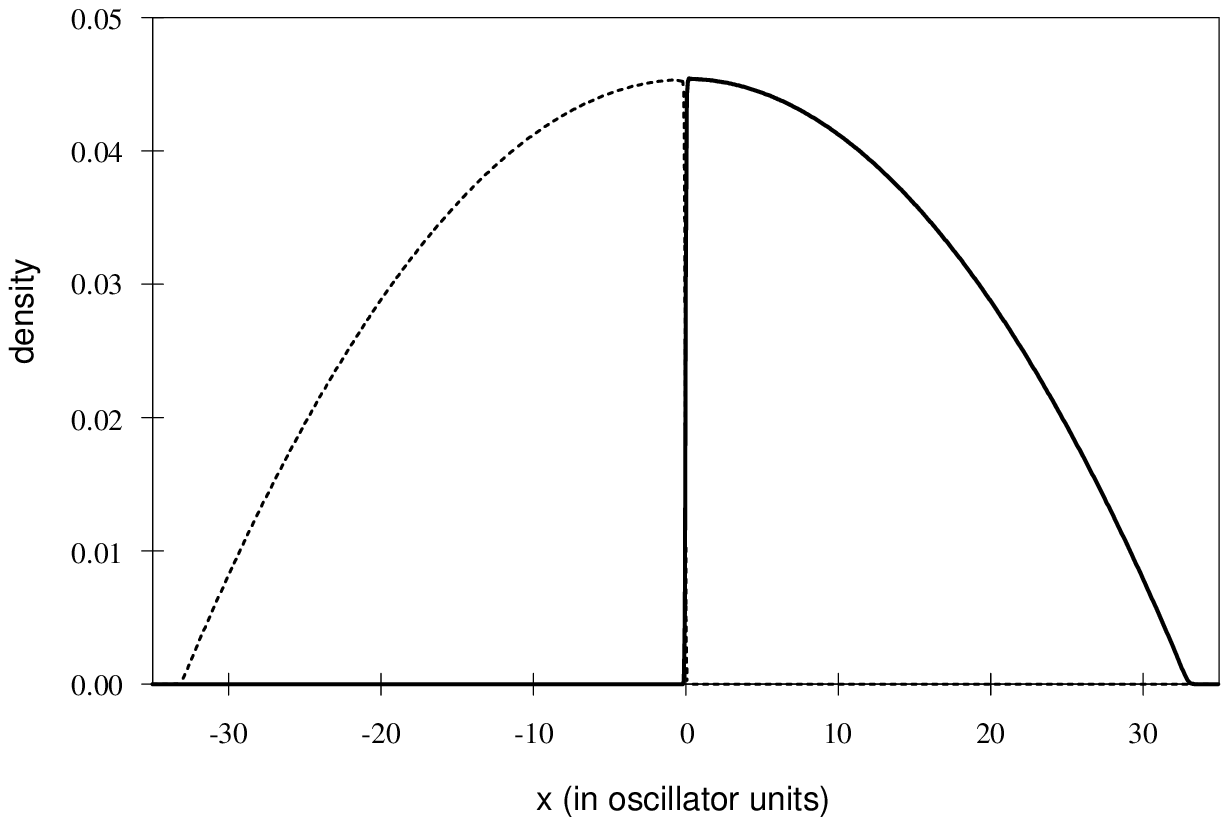}}
\caption{Two types of solutions without overlap: symmetric and
asymmetric cases.  Plotted are densities for the first and second
components as solid and dashed curves.  For both panels, the ratio of
the scattering lengths is $U_{11}:U_{12}:U_{22}=1:1.52:1.01$ and
$N_{1}=N_{2}$.  The ratio of the energy of the asymmetric case to the
symmetric one is $0.8$.}
\label{nooverlap}
\end{figure}

Eigenstates of the binary-condensate system that break the symmetry of
the trapping potential exist.  In 1D, a solution of this kind is given
by TF parabolas that are stuck together.  An example is shown in
Fig.~\ref{nooverlap}.  In this case one can derive equations for $\mu
_{1}$ and $\mu _{2}$ in the same way as in Sec.~\ref{1D case},
integrating the densities and substituting the result into the
normalization conditions.  The generalization to the dimensions higher
than one may be only obtained if the separation surface (which reduces
in 1D to a single point) is simple.  In 1D, the corresponding coupled
equations
take the form
\begin{equation}
\frac{(2\mu_1)^{3/2}}{3\sqrt{m\omega^2}} - \mu _{1}R + \frac{1}{6}
m\omega^2 R^3 = U_{11}\, N_1
\end{equation}
\begin{equation}
\frac{(2\mu_2)^{3/2}}{3\sqrt{m\omega^2}} + \mu _{2}R - \frac{1}{6}
m\omega^2 R^3 = U_{22}\, N_2
\end{equation}
\begin{equation}
{\cal E} = ( \frac{(2\mu_1)^{5/2}}{5\sqrt{m\omega^2}} - \mu _{1}^2
R + \frac{1}{20}(m\omega^2)^2R^5)/U_{11} + (
\frac{(2\mu_1)^{5/2}}{5\sqrt{m\omega^2}} + \mu _{2}^2R -
\frac{1}{20}(m\omega^2)^2R^5)/U_{22}
\end{equation}

We have checked numerically that this solution cannot give rise to
a minimum of the free energy, hence, within the framework of the
FT approximation, the ground state cannot be the one with broken
symmetry. The difference in the energy between symmetric and
asymmetric cases is usually very small making them almost
degenerate. The degeneracy is exact in the limit when $U_{11}
\rightarrow U_{22}$ and is removed by the kinetic energy.

\section{The role of kinetic energy}

In this section we present results of our studies of the
contribution of the kinetic energy on the total energy and on the
functional dependence of the ground state of the binary mixtures
of the BEC. For a single condensate, in the regime of the validity
of TF approximation, the kinetic energy creates a healing length,
$\xi$, in the region where the condensate wavefunction tends to
zero.  This healing length is of the order of $(8\pi n a)^{-1/2}$
\cite{stringari}, where $a$ is a scattering length and $n$ is an
average density of the condensate.  For the binary mixtures of
BECs there are two length scales that can be defined in order to
characterize two kinds of boundary regions.  One is the ordinary
healing length of single condensate and refers to the healing of
the wavefunction outside region of the coupled wavefunctions.  But
for mixtures another region between the two components exist and a
penetration depth, $\chi$, as a length scale over which two
components overlap.  The penetration depth is a function of $\det
U$.  For $\det U < 0$ the lowest energy state consists of
partially overlapping wavefunctions; hence the penetration depth
is of order of the size of the condensate.  With decreasing $\det
U$ the penetration depth becomes smaller and goes to zero in the
limit of strong repulsion as $\det U \rightarrow -\infty$. At the
same time, the contribution of the kinetic energy to the total
energy becomes more important, in spite of the shrinking overlap
region. This situation is illustrated in Fig.~\ref{u12}. The
energy of the lowest eigenstate is plotted in the symmetric and
asymmetric classes as a function of $\det U/(U_{11}U_{22})$.
Fig.~\ref{u12} Only one curve is plotted for $\det U > 0$ where
the contribution of the kinetic energy is negligible and
Thomas-Fermi approximation gives an excellent prediction for both
the value of the ground state energy and its wavefunction.  As the
value of $\det U$ becomes negative, the lowest energy state within
a TF approximation consists of two separated components and the
energy does not depend on $U_{12}$ and is plotted as a horizontal
dashed line in Fig.~\ref{u12}.  $\det U < 0$ the contribution of
the kinetic energy is substantial and it is larger for the
symmetric case which has two interfaces between phases. The
asymmetric configuration has only one interface in the lowest
energy state. Hence, the ground state looses symmetry of the trap.

\begin{figure}[ht]
\centerline{\epsfxsize=4.25in\epsfbox{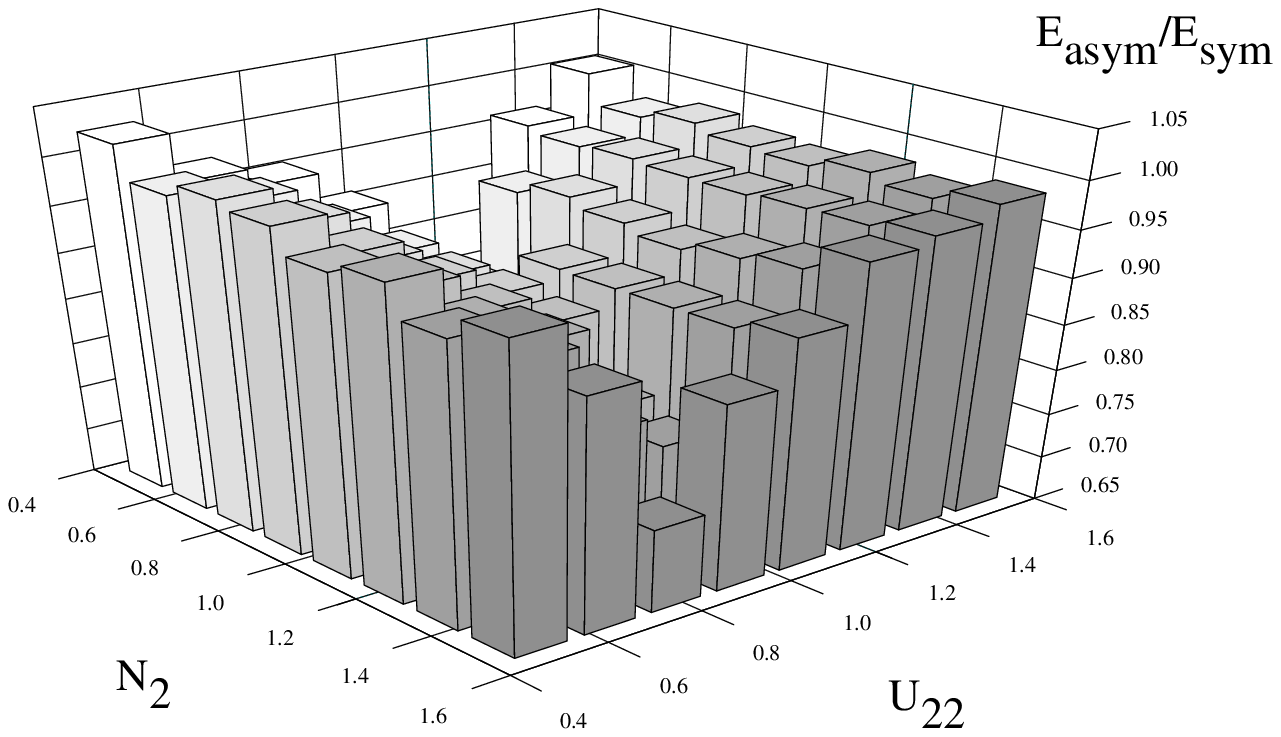}}
\caption{Ratio of the total energy of the asymmetric case to the
corresponding value in the symmetric phase vs.  $U_{22}$ and $N_{2}$
scaled by certain initial values.  The initial values are the same as
in Fig.~\ref{nooverlap} (i.e., $U_{11}:U_{12}:U_{22}=1:1.52:1.01$ and
$N_{1}=N_{2}$).  Note the deep valley near $U_{11}=U_{22}$.}
\label{N2-u2}
\end{figure}

In order to search for stable symmetry-breaking solutions we kept
$N_{1}$,$U_{11}$ and $U_{12}$ constant and varied $N_{2}$ and
$U_{22}$.  Fig.~\ref{N2-u2} plots the ratio of the total energy in the
asymmetric case to the energy of the symmetric one as a function of
these two variables.  Almost all solutions are symmetry-breaking ones
and a trough is formed near $U_{22}=U_{11}$.  The trough is an optimal
region for finding symmetry-breaking solutions.

\section{Finite Interaction Range}

We have so far considered the mean-field description of a BEC mixture
assuming a zero-range delta-function pseudopotential.  It is of
interest to consider the effects of a finite-range interatomic
interaction on the ground-state structure in the two-component
condensate.  We introduce a pseudopotential in the form of a
normalized Gaussian with a finite width (i.e., range) which recovers
the zero-range limit result as the range vanishes.  We search for
changes in the structure of the ground state of the two-component
system as the range of the {\em intercomponent} interactions only are
varied, keeping the delta-function pseudopotential for the
self-interactions.  The results displayed below were obtained by means
of direct numerical simulations, not the FT approximation.  In 1D, the
intercomponent-interaction terms in Eqs.~(\ref{GP1}) $U_{12}|\psi
_{i}(x,t)|^{2}$ are replaced by a nonlocal expression,
\begin{equation}
U_{12} \frac{1}{\sqrt{2\pi d^2}} \int_{-\infty }^{+\infty }dy\
\exp (\frac{ -(x-y)^{2}}{2d^{2}}) \,|\psi _{i}(y,t)|^{2}{\rm \,},
\label{gauss}
\end{equation}

where $d$ is the interaction range.  The Gaussian form was chosen
to model a finite-range potential solely for its simplicity (see
also Ref.~\cite{Salasnich} for the role of a finite interaction
range in attractive single-component BEC within this model).  This
family of finite-range potentials has a constant scattering length
within the first Born approximation.  All the cases that we
investigate below correspond to configurations with {\em
separated} wavefunctions in the usual TF limit (see
Sec.~\ref{separated}).

\subsection{Weak separation $(U_{11}U_{22}\leq \left( U_{12}\right)
^{2}) $}

\begin{figure}[ht]
\centerline{\epsfxsize=4.25in\epsfbox{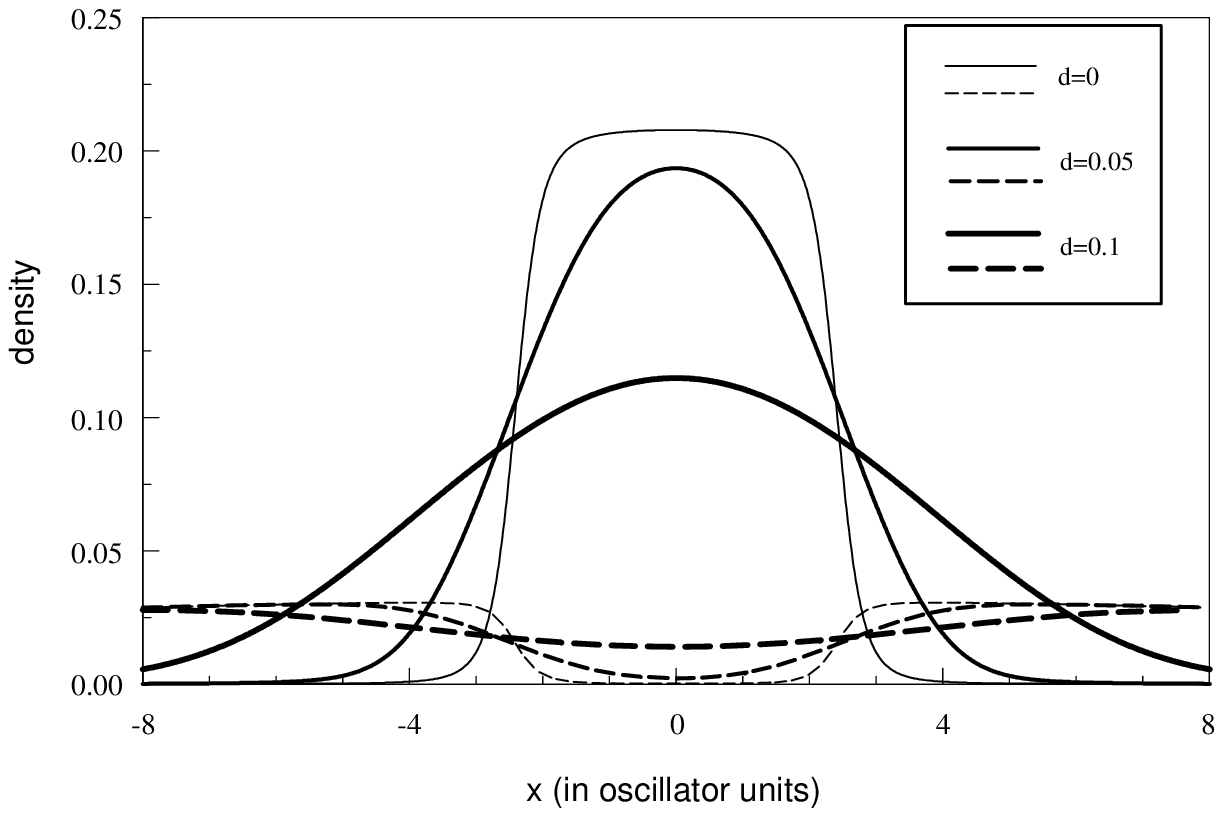}}
\caption{Dependence of the density distributions of a two-component
BEC in weakly segregated phases on the range of the intercomponent
interaction $d$.  Here $U_{12} = 1.02\,U_{11} = 1.02\,U_{22}$ and
$N_{2}/N_{1}=0.15$.  The effect of attenuation of the interaction with
growth of the range is illustrated for three values of $d$.}
\label{soft}
\end{figure}

Here we discuss case 2 of Table 1.  Fig.~\ref{soft} shows the overlap
region between the wavefunctions of the two components; this overlap
region grows as the interaction range increases.  Starting from
relatively well separated phases, we end up with a complete overlap of
the two components (i.e., the component located initially outside the
narrow-width component finally penetrates it).  The intercomponent
interaction parameters are chosen as $U_{12} = 1.02\,U_{11} =
1.02\,U_{22}$, which places this case not far from the boundary
between the cases of separated and overlapping phases.  In other
words, increasing the interaction range forces a transition from
separated phases to penetrating ones.  Within the zero-range model,
this would correspond to a transition from $U_{12} \geq
\sqrt{U_{11}U_{22}}$ to $U_{12}\leq \sqrt{U_{11}U_{22}}$, i.e., to
attenuation of the interaction.  A simple argument justifies this
conclusion.  Suppose the interaction range is so large that the
long-range potential varies only slightly over the extent of the
interface between the nearly separated components (where the
intercomponent interaction is important).  Then, the interaction terms
of Eq.~(\ref{gauss}) would be approximately constant, producing only a
shift in the total energy of the mixture, but not affecting the shape
of the two wavefunctions.  When $d$ becomes comparable to the
penetration depth (see also Ref.~\cite{Ao}), phase separation is
reduced.  This is clearly illustrated in Fig.~\ref{soft}.

\begin{figure}[ht]
\centerline{\epsfxsize=4.25in\epsfbox{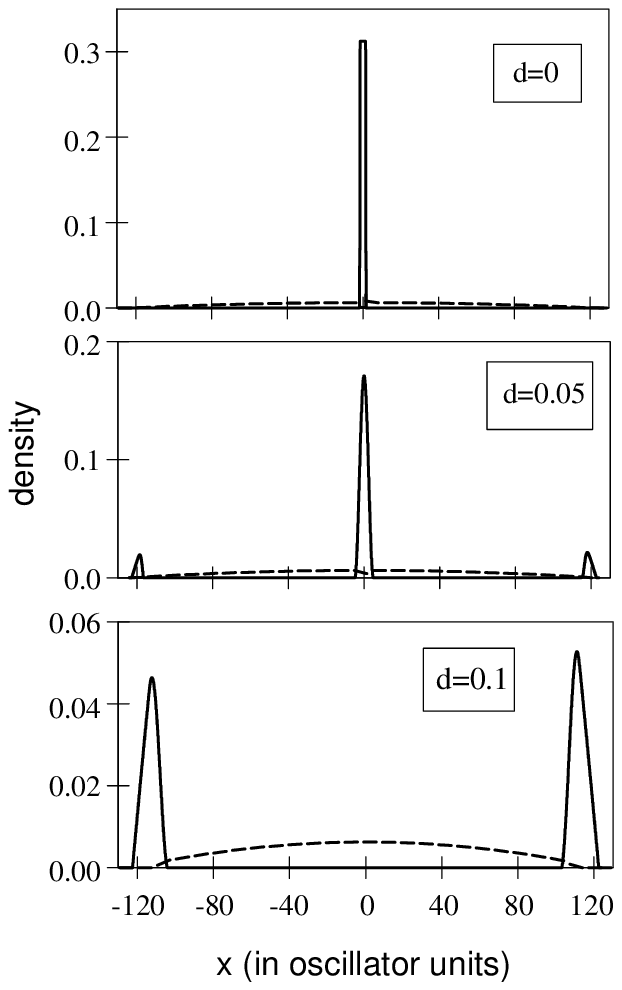}}
\caption{Interaction-range induced change of the ground-state
density distribution of a two-component BEC in weakly segregated
phases, but with a large disparity in the number of atoms in the
components. The parameters are $U_{12}=1.01\,U_{11}=1.01\,U_{22}$
and $N_{2}/N_{1}=0.02$.} \label{topol}
\end{figure}

Although the parameters in the 1D solution are not directly relevant
to an experimental situation, we decreased the range of the
intercomponent interactions so that they are in a realistic range of
values.  From the aforementioned arguments, we see that in order to
observe a difference from the zero-range case, the interaction range
should be comparable to the penetration depth.  In order for the
interaction range to be small and yet correspond to the boundary
between overlapping and segregated phases, a large disparity in the
number of atoms in the two components is required.  This situation is
depicted in Fig.~\ref {topol}, where $U_{12} = 1.01\,U_{11} =
1.01\,U_{22}$, but $N_{1} \gg N_{2}$ .  We observe a qualitative
change in the ground-state solution: the component that initially was
at the center of the trap moves to its outskirts as the interaction
range grows.  Although $d = 0.1 \, \sqrt{\hbar/m \omega}$ (bottom
frame in Fig.~\ref{topol}) would usually correspond to several tens of
nanometers, we argue that one can optimize the sensitivity to $d$ by
considering the regime of parameters near the boundary between the
penetrating and segregated phases.  The possibility of manipulating
the strength of atomic collisions is not excluded, as Feshbach
resonances have been observed in BEC samples \cite{feshbach}, and
several other proposals in this respect have been put forward
\cite{modscatt,goldstein}.  Using such techniques one could vary the
intercomponent scattering length in order to scan the region near
$\det U=0$.  Then, by comparing the measured structure of the mixture
to the predictions of the two theoretical models (i.e., the ones with
zero and finite interaction range) one could determine the effective
range of interactions (which is a parameter in the latter model). 
Thus, one can probe the microscopic parameter $d$ via a magnified
effect such as the qualitative change of the condensate structure from
to separated phase to penetrating phase.

\subsection{Strong separation $(U_{11}U_{22}\ll \left(U_{12}\right)
^{2})$}

\begin{figure}[ht]
\centerline{\epsfxsize=4.25in\epsfbox{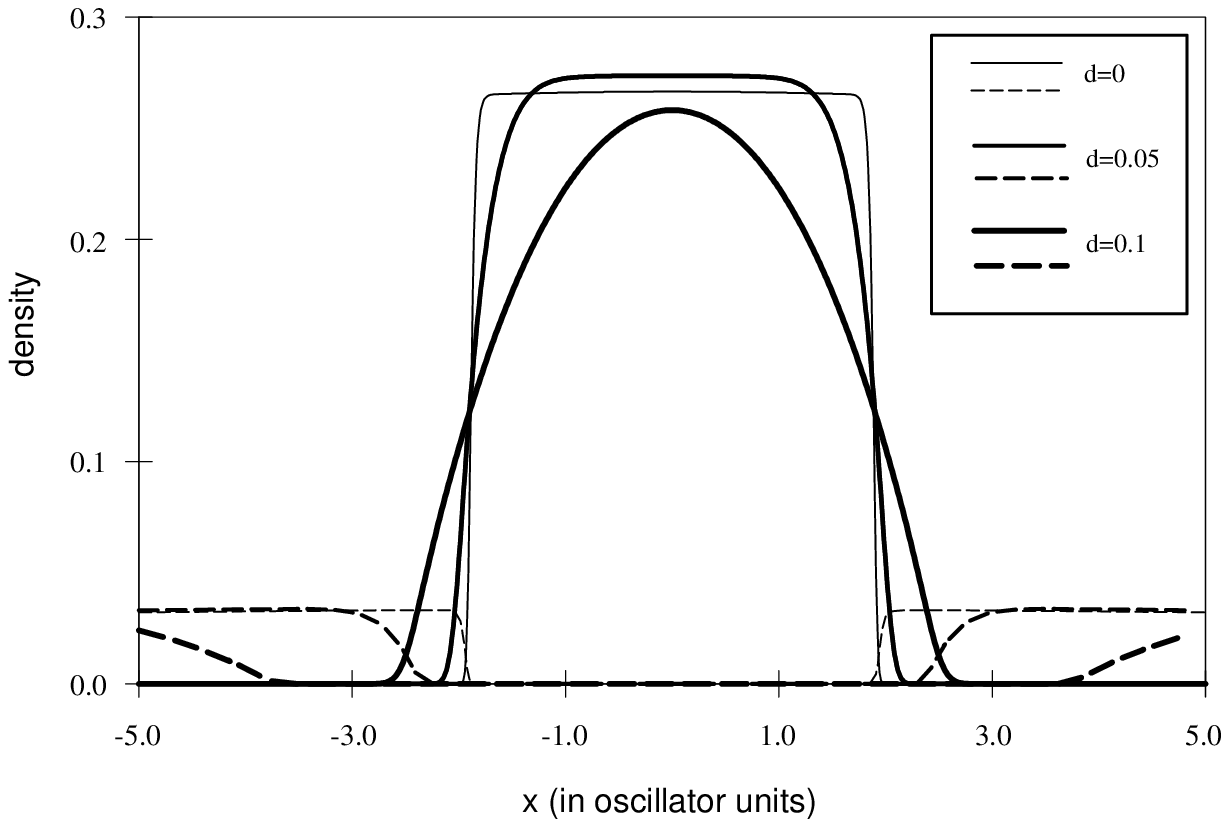}}
\caption{Enhanced mutual repulsion as an effect of a finite
interaction range between strongly separated phases. Here
$U_{11}:U_{12}:U_{22}=1:7:1$ and $N_{2}/N_{1}=8$.} \label{sharp}
\end{figure}

In the preceding section we demonstrated the effective attenuation of
the mean-field repulsive interaction between two components of a BEC
due to an increase in the range of the interaction for the case when
$U_{11}U_{22} \leq ( U_{12})^{2}$.  Now we turn to the case when the
parameters of the BEC mixture are far from threshold for the onset of
the penetrating phase.  In Fig.~\ref{sharp} is for parameters $U_{12}
= 7\,U_{11} = 7\,U_{22}$, hence the interface between the two
components is very sharp.  Since now it is much easier to match the
range of interactions with the penetration depth, one might expect
that effects similar to those described above will appear at even
smaller values of $d$.  However, this is not the case.  The mutual
repulsion of the components remains very strong even if reduced by a
finite interaction range.  As the interaction range increases, the two
components tend to move apart, yielding two completely separated
phases.

\section{Summary and Conclusion}

\begin{figure}[ht]
\centerline{\epsfxsize=4.25in\epsfbox{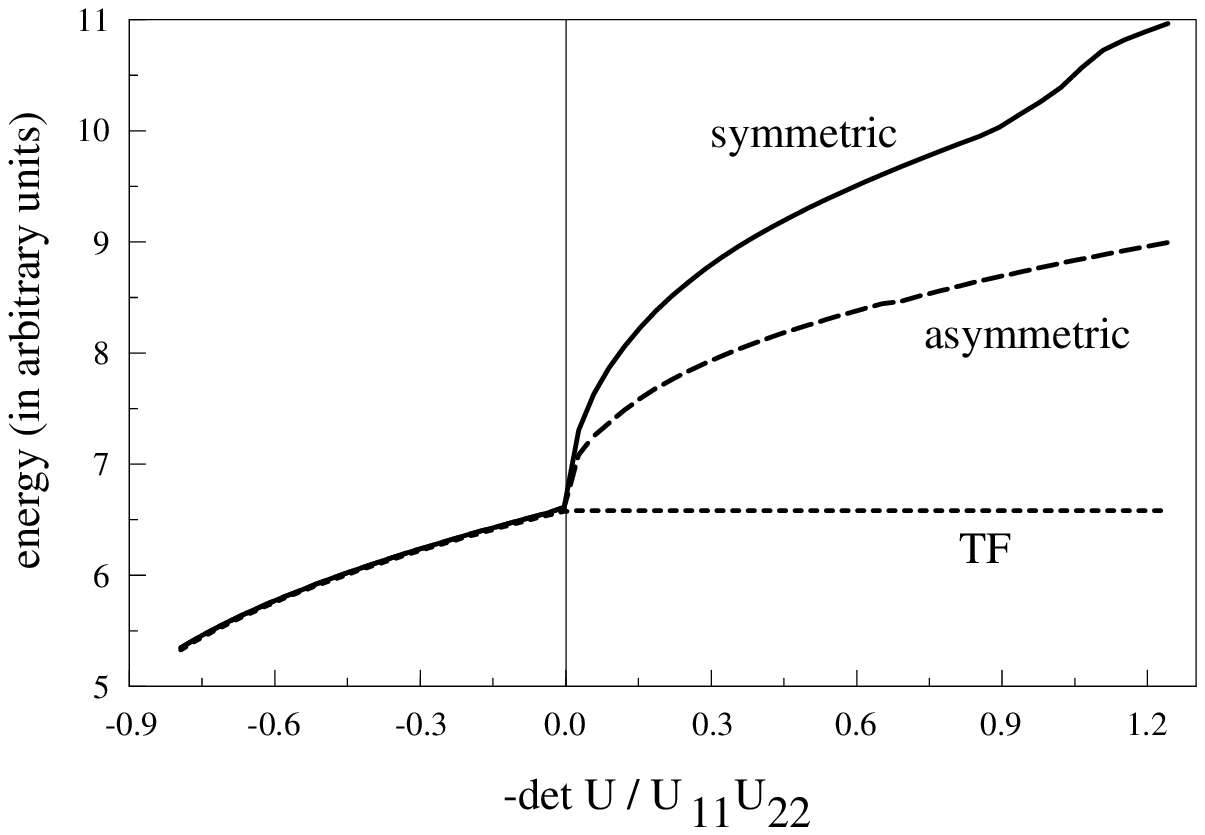}} \caption{Total
energy vs.~the dimensionless quantity $\det U/(U_{11}U_{22})$ for
the symmetric and asymmetric cases (numerical simulation) and the
TF predicion.  Here $N_{1}=N_{2}$, $U_{11}$ and $U_{22}$ are kept
constant (the same as in Fig.~\ref{nooverlap}) whereas $U_{12}$ is
varied.  For arguments smaller than $0$ the three curves are
indistinguishable.} \label{u12}
\end{figure}

We have presented a detailed classification of stable solutions for
binary mixtures of dilute atomic condensates.  The analysis is
particularly simple within the Thomas-Fermi approximation.  Within
this approximation one can distinguish two general classes of ground
state for two component condensate mixtures: unseparated ones with an
overlap region (both component wavefunctions are simultaneously
nonzero) and separated ones not containing an overlap region (except
for the tail penetration).  The latter contains also solutions that do
not posses the symmetry of the trapping potential.  Components are
separated if $\det U \equiv U_{11}U_{22} < U_{12}^2 < 0$ and they
overlap if $\det U > 0$.  The predictions from the TF approximation
become ambiguous in the region of parameters where phase separated
solutions which break the symmetry of the trap are energy-degenerate
with the phase-separated solutions preserving the symmetry.  In this
case, it is crucial to include the contribution of the kinetic energy
operator and of the mutual interaction energy in determining the
structure of the ground state geometry, which tend to favour the
asymmetric solutions.  The physical reason for this is probably a
smaller interface region in the asymmetric solutions.  It is these
interface regions which contribute most to both the kinetic and mutual
interaction energies.  We have carried out our numerical calculations
in 1D, but our conclusions should be valid in two and three dimensions
as well.  The condensates, if not overlapping, should have a
propensity towards states with minimal interface surface area.  In the
$\det U < 0$ case, with kinetic energy included, we further recognize
a weak separation regime ($\det U \leq 0$) and strong separation
regime ($\det U \ll 0$) when a penetration depth goes to zero. 
Contribution of the kinetic energy to the total energy increases with
decreasing $\det U$ (for negative $\det U$), in spite of the
decreasing interface size (see Fig.~(\ref{u12}).  This is due to the
increasing importance of the $U_{12}$ term that gives the
cross-interaction energy of atoms from different components.

Since the size of the overlap region can be very small (smaller than a
single condensate healing length), it is of interest to investigate
the possible impact of a small, but non-zero, interaction range in a
binary condensate.  We have developed a model of finite range
potential by introducing a pseudopotential in the form of a normalized
Gaussian with finite width.  We identified two distinctly different
cases.  In one case, that of significant overlap, the finite range
tends to increase the penetration over the delta function interaction,
and in the other, that of strong separation, the finite range leads to
a trough between the two condensates.

\bigskip
We thank M. Gajda and J. Mostowski for stimulating discussions.
This work was supported in part by the US-Israel Binational
Science Foundation and the James Franck Binational German-Israel
Program in Laser-Matter Interaction (Y.B.B.).  K.R. and K.G.
acknowledge the support of the subsidy from the Foundation for
Polish Science and of the Polish KBN Grant no 2 P03B 057 15. Part of
the results have been obtained using computers at the
Interdisciplinary Center for Mathematical and Computational
Modeling (ICM) at Warsaw University.

\end{document}